# Three individual two-axis control of singlet-triplet qubits in a micromagnet integrated quantum dot array

Wonjin Jang[1†], Min-Kyun Cho[1†], Jehyun Kim[1], Hwanchul Chung[2], Vladimir Umansky[3], and Dohun Kim[1]*

[1]*Department of Physics and Astronomy, and Institute of Applied Physics, Seoul National University, Seoul 08826, Korea*

[2] *Department of Physics, Pusan National University, Busan 46241, Korea*

[3]*Braun Center for Submicron Research, Department of Condensed Matter Physics, Weizmann Institute of Science, Rehovot 76100, Israel*

[†]*These authors contributed equally to this work*

*Corresponding author: dohunkim@snu.ac.kr*

**Abstract**

We report individual confinement and two-axis qubit operations of two electron spin qubits in GaAs gate-defined sextuple quantum dot array with integrated micro-magnet. As a first step toward multiple qubit operations, we demonstrate coherent manipulations of three singlet-triplet qubits showing underdamped Larmor and Ramsey oscillations in all double dot sites. We provide an accurate measure of site-dependent field gradients and rms electric and magnetic noise, and we discuss the adequacy of simple rectangular micro-magnet for practical use in multiple quantum dot arrays. We also discuss current limitations and possible strategies for realizing simultaneous multi-qubit operations in extended linear arrays.

Fabrication of large array of qubits and demonstration of coherent multi-qubit operations are necessary steps for realizing scalable quantum processing unit[1–4]. Harnessing spin degree of freedom in gate defined quantum dots (QDs) has attracted significant interests [5–14] owing to spin's long coherence time in solid state system and potential scalability including well-established fabrication technology[2,15,16]. Depending on the degree of spin-charge mixing, elementary quantum operations of Loss-DiVincenzo (single electron)[12,14,17,18], singlet-triplet ($ST_0$, two electron)[5,10,11,19–21], and various three electron spin qubits[22–24] have been demonstrated. Electron loading and charge state manipulation are also performed in ~ 10 coupled linear dots[3,25] and ~ 4 site two dimensional arrays[1,4]. Moreover, QD Hamiltonian parameters have shown to be widely tunable[19,26–28]. With possibility of efficient electrical control, QD systems are also emerging as promising quantum simulators[26,29,30].

Toward realizing multi-qubit operations in semiconductor quantum chips, coherent qubit addressing beyond ground state property tuning is important. In this work, we demonstrate individual operations of three $ST_0$ qubits formed in different double dot sites in a sextuple linear quantum dot array with rf-single-electron transistor (rf-set) sensors. Following pioneering works such as recent demonstration of four individual Loss-DiVincenzo qubit operations in GaAs[31], we employed proximal rectangular micro-magnet enabling sizable magnetic field gradient[32] and two-axis control for all $ST_0$ qubits in our device. This work addresses spatial range of experimentally usable field gradient by simple micro-magnet using qubit oscillations as sensitive magnetic probes, and we show that magnitude of electric and magnetic noise of all qubits are comparable to the previous works. Moreover, we discuss examples of adverse effects of surface micro-magnet structure fabricated on the device without interposing dielectrics, which calls for further optimization of sensor and magnet position to enable simultaneous qubit and sensor operations in similar device geometries.

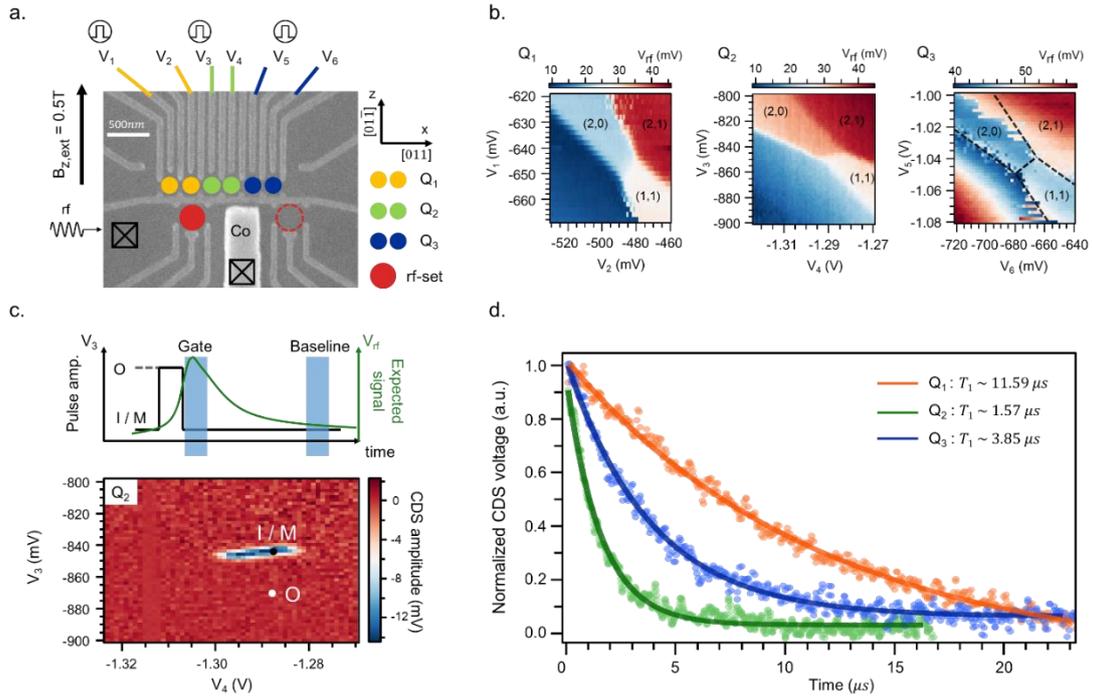

**Figure 1.**

Figure 1a shows the scanning electron microscope image of the QD array device identical to the sample we measured. The device consists of linear sextuple QD array with rf-set sensors (only rf-set1 is operated here) metallized with Au/Ti gates on top of a GaAs/AlGaAs heterostructure, where a 2D electron gas (2DEG) is formed approximately 70 nm below the surface. A 250 nm -thick rectangular Co micro-magnet with large shape anisotropy was deposited to generate stable site dependent magnetic field difference $\Delta B_z$ for $ST_0$ qubit operation. The device was placed on a plate in a dilution refrigerator at ~20 mK and in-plane magnetic field $B_{z,ext}$ of 500 mT was applied. The electron temperature, estimated with Coulomb blockade thermometry[33,34], is approximately 230 mK [35].

We independently operate and readout three $ST_0$ qubits ($Q_1$ to $Q_3$), where we monitor the rf-reflectance of the rf-set1 for all qubits. We set integration time of the rf-demodulator to 30 ns to monitor fast transient signals. Three sets of double QD charge stability diagrams are shown in Fig. 1b with respective regions of two electron occupancies with inter-dot tunnel coupling $t_c$ tuned above $8~\text{GHz}\cdot\text{h}$, where h is the Planck's constant. We stress here that these charge stability measurements are independently performed; while $Q_1$ to $Q_3$ are sequentially tuned, charge occupancies other than dots under investigation are not strictly examined. More systematic tuning such as virtual gate-based '$n+1$' method[3,25,26] or fast raster scanning[3,36,37] can be applied in the future to set, for example, strict one electron per dot in all QD sites. Including improvement in the calibration speed, we discuss current limitations in the results section.

The spin-to-charge conversion is performed using Pauli spin blockade (PSB) with fast correlated double sampling (CDS) technique[35,38]. Figure 1c top panel shows data sampling scheme, where ~200 ns long gate and baseline signals are subtracted and averaged 4096 times. Figure 1c bottom panel shows an example CDS map of $Q_2$, where non-zero triplet return probability induced by fast control pulse is reflected by non-zero signal near (2,0) to (1,1) charge transition. In this PSB readout position, the triplet state is mapped to (1,1) charge occupancy before it is relaxed to singlet (2,0) state with relaxation time $T_1$, producing signal difference between gate and baseline signal.

Figure 1d shows $T_1$ times at PSB readout detuning positions for $Q_1$ to $Q_3$. Ranging from $T_1 \sim 1\mu s$ ($Q_2$, largest $\Delta B_z$) to $T_1 \sim 10\mu s$ ($Q_3$, smallest $\Delta B_z$), the trend shows that the $T_1$ in our device is mainly limited by the $\Delta B_z$ induced relaxation[19,39,40]. Fast relaxation at I/M positions prevents performing high fidelity single-shot readout using conventional PSB readout especially in the large $\Delta B_z$ regime. Recent works study various alternative schemes to overcome shortcomings of the PSB readout using electron latching[35,38,40,44]. We expect that latched readout can be straightforwardly used for $Q_1$ and $Q_3$ while reading out $Q_2$ isolated from the reservoir from both sides needs modification of device design. Keeping experimental simplicity, we report all the qubit operation data in box-car averaged manner in the current work using conventional PSB readout and leave optimization of single-shot schemes as future works.

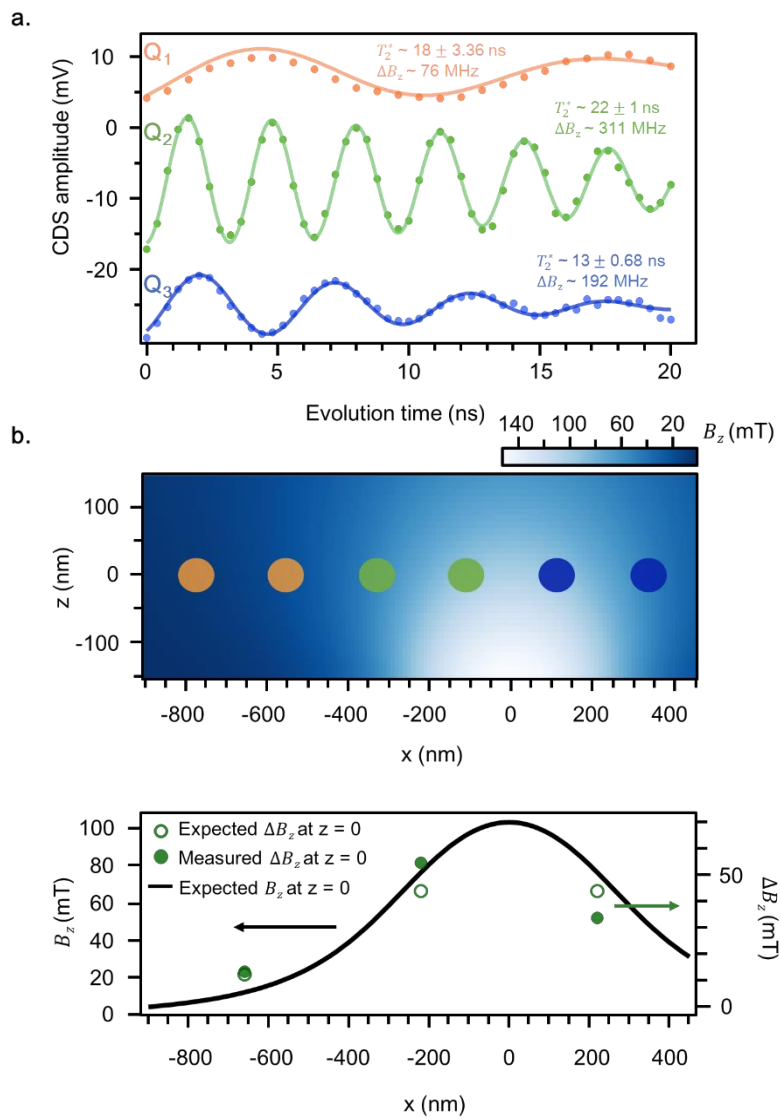

**Figure 2.**

We demonstrate three Larmor oscillation measurements by applying the I – O – M pulse sequence with repetition rate of 25 kHz. The I/M positions reside in the PSB to discriminate the different spin states and to enable the initialization by waiting for period 40 $\mu s$ longer than the $T_1$ of all qubits. Varying evolution time $\tau$ at the point O where exchange energy $J\sim 0$ [5],

$\Delta B_z$ induces ST$_0$ oscillations corresponding to rotation around x-axis of the each qubit's Bloch sphere, where the CDS settings are adjusted to yield the CDS voltage proportional to the T$_0$ probability for all qubits throughout Fig. 2, and Fig. 3. As shown in Fig. 2a, ST$_0$ oscillation of all qubits as a function of $\tau$ exhibits underdamped probability oscillation with distinct frequency set by site dependent $\Delta B_z$. The measurement time for each trace (average of 10 repeated identical trace) is about 1 min, longer than typical nuclear fluctuation time scale, and we observe ensemble averaged, inhomogeneous coherence time $T_2^*$ on the order of 10 ns limited by nuclear field fluctuation, consistent with existing reports[5,19,45].

Figure 2b shows simulation of two-dimensional map of magnetic field produced by micro-magnet using the boundary integral method with RADIA software[46,47]. The z-component of the magnetic field, $B_z$ varies from 7.4 mT to 96.1 mT depending on the relative position of the micro-magnet according to the simulation, and the spatial field difference along each qubit's constituent quantum dot distance is calculated to be 12.3 mT (~ 70 MHz·h/(g*$\mu_B$)), 43.9 mT (~ 250 MHz·h/(g*$\mu_B$)), 43.9 mT (~250 MHz·h/(g*$\mu_B$)) for Q$_1$, Q$_2$ and Q$_3$ respectively (Fig. 2b lower panel), where $g^*$ is the electron g-factor in GaAs and $\mu_B$ is the Bohr magneton. The calculated $\Delta B_z$ agrees well with the experimental observation shown in Fig. 2a, and we ascribe the frequency discrepancy between Q$_2$ and Q$_3$ to the misalignment of the micro-magnet from the intended position or unintended asymmetry of the magnet shape occurred in the fabrication process. Confirming the validity of the simulation, we expect 4 ~ 5 ST$_0$ qubits spanning over 1 $\mu m$ array length can show underdamped x-axis rotation operations in this ensemble averaged $T_2^*$ time scale with the current micro-magnet design. Possible routes for improvement include fast Hamiltonian estimation to extend $T_2^*$ time scale[48,49] and periodic deposition of the magnet structure with intermediate rf sensors to extend number of qubits.

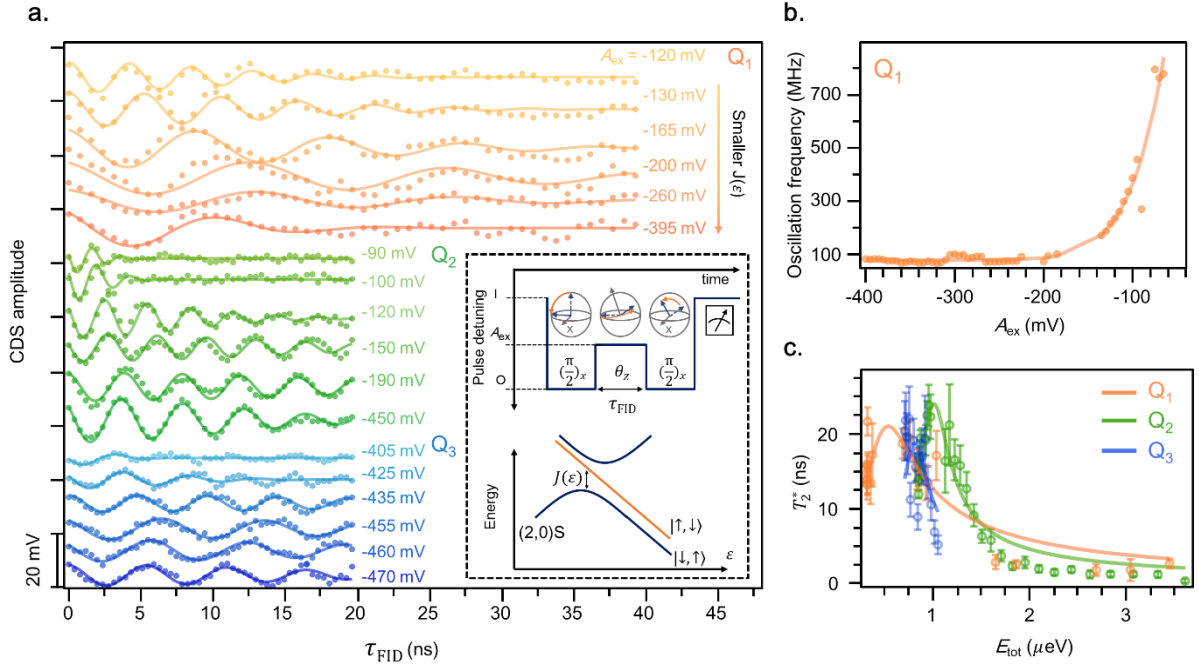

**Figure 3.**

We turn to discuss two-axis control of three $ST_0$ qubits. Using a typical Ramsey pulse sequence I– O ($\pi/2$) – $A_{ex}$ – O ($\pi/2$)– M (Fig. 3a, upper panel in the inset), Fig. 3a shows the coherent manipulation of $Q_1$ to $Q_3$ as a function of detuning pulse amplitude $A_{ex}$ and delay time $\tau_{FID}$ recorded with a single rf-set. The plots show coherent quantum oscillation in all qubits as well as continuous evolution of rotation axis on the Bloch sphere from z- to x- axis as $A_{ex}$ is varied over different regimes, where $T_2^*$ is limited by the charge noise for $J(\varepsilon) > \Delta B_z$ or by fluctuations in $\Delta B_z$ for $J(\varepsilon) \sim 0$. The size of the $J(\varepsilon)$ can be evaluated by assuming the exponential dependence of $J(\varepsilon)$ on the detuning $\varepsilon$ [21,50] (see Fig. 3a inset, bottom panel), and by fitting the free induction decay (FID) oscillation frequency f($\varepsilon$) to $f(\varepsilon) = \sqrt{(\Delta B_z)^2 + J(\varepsilon)^2}/h$ (Fig. 3b). Then, by investigating the $T_2^*$ dependence on the qubit energy splitting $E_{tot}(\varepsilon) = h \cdot f(\varepsilon) = \sqrt{(\Delta B_z)^2 + J(\varepsilon)^2}$, we obtain the root-mean-squared noise amplitudes of the magnetic field $\delta B$ and the charge fluctuation $\delta \varepsilon$ of the three qubits[21]. Assuming the gaussian decay of the FID curves, we first acquire the $T_2^*$ along the $A_{ex}$ where the $A_{ex}$ can be converted into the $E_{tot}$ according to the fitting shown in Fig. 3b. Considering charge and magnetic field fluctuations up to the first order, the total noise amplitude of the $E_{tot}$ can be modeled as $\delta E_{tot}(\varepsilon) = \sqrt{(\frac{\Delta B_z}{E_{tot}(\varepsilon)} \delta B)^2 + (\frac{J(\varepsilon)}{E_{tot}(\varepsilon)} \frac{dJ(\varepsilon)}{d\varepsilon} \delta \varepsilon)^2}$, where $T_2^* = \frac{\sqrt{2}\hbar}{\delta E_{tot}}$ holds. In Fig. 3c the $T_2^*$ vs. $E_{tot}$ plots of the three qubits are shown, where the fitting yields the magnetic field noise amplitude of ($52.0 \pm 3.0\,neV$, $39.2 \pm 2.2\,neV$, $50.1 \pm 5.79\,neV$), and the charge noise amplitude of ($6.3 \pm 0.6\,\mu eV$, $5.3 \pm 0.4\,\mu eV$, $9.5 \pm 2.5\,\mu eV$) for ($Q_1$, $Q_2$, $Q_3$). The obtained $\delta\varepsilon$ are similar to the values in the precedent works[21,41], and the $\delta B$ are consistent with the $T_2^* = \sqrt{2}\hbar/\delta B$ shown in Fig. 2b where $J(\varepsilon) \sim 0$.

We note that the high frequency wiring for $Q_3$ shows attenuation heavier than the other lines (- 20 dB Vs. - 27 dB), so that the detuning energy modulation amplitude for $Q_3$ is somewhat limited. Moreover, since only one high-frequency line per qubit is available in the current experiment (see Fig. 1a), we were unable to perform multidirectional pulsing. Signal to noise ratio for $Q_3$ measurement is also lower than the other qubit measurements as the $Q_3$ is farther away from the sensor. The device and rf-setup indeed support dual rf-reflectometry, but we operate only rf-set1 for the following reason. While depositing thick micro-magnet directly on the device surface without insulating interpose layer simplifies fabrication steps, we find this step unintentionally leads to carrier depletion under the magnet area, likely due to interface strain[51] or plasma treatment before deposition[52], for all the devices we tested ( > 10 devices in two different batches with the same design). Along with carrier depletion under nearby gate electrodes, the net effect blocks current path to rf-ohmic when both rf-set gate voltage sets are sufficiently negative, which precludes simultaneous use of two rf-sensors tuned to maximum charge sensitivity in our device. In the Supplementary Information, we show the evidence of this effect measured by dc transport along with respective rf-set sensor tuning. In the future device design, we intend to use the magnet as gate electrode, as opposed to the present floating magnet, to investigate if carrier can be accumulated near magnet region with proper tuning. Overall, while the experiment shows successful initial demonstration for multiple qubit operation in QDs by straightforward extension of quantum measurement schemes developed in the past[5,35,38,43,53], fuller control and measurement of qubits at least require larger number of high-frequency line setup and further optimization of magnet position or gate electrode to magnet clearance.

In conclusion, we have shown three independent $ST_0$ qubit operations in a linear sextuple quantum dot array in GaAs. The main achievement toward fuller realization of three qubit quantum processor is the demonstration of three underdamped quantum oscillations enabling independent two-axis control on the respective qubit's Bloch sphere. Through $ST_0$ oscillation measurements, site-dependent $\Delta B_z$ was analyzed where we show a simple rectangular micro-magnet proximal to a quantum dot array in the same gate layer can produce sizeable $\Delta B_z$ for over $\sim \mu m$ spatial range with sufficiently simple fabrication step without adding too much electric and magnetic noise. Overcoming current limitations in our device including application of single-shot read out dealing with short $T_1$ times and adding gating capability for micro-magnets to enable simultaneous rf-set readouts, we plan to study simultaneous three qubit operations and pair-wise two qubit interactions in the improved device designs.

**Acknowledgements**

This work was supported by Samsung Science and Technology Foundation under Project Number SSTF-BA1502-03. Cryogenic measurement used equipment supported by the National Research Foundation of Korea (NRF) Grant funded by the Korean Government (MSIT) (No.2019M3E4A1080144, No.2019M3E4A1080145) and the Creative-Pioneering Researchers Program through Seoul National University (SNU).

**Data Availability Statement**

The data that support the findings of this study are available from the corresponding author upon reasonable request

**Figure captions**

**Figure 1**. **Measurement setup for three singlet-triplet qubit operation. a.** Scanning electron microscope image of the device identical to the sample utilized for the experiment. 3 different singlet-triplet ($ST_0$) qubits are defined in the 6-quantum dot (QD) array, and gates $V_1$, $V_3$ and $V_5$ are used for fast pulsing. Micro-magnet generates the spatial magnetic field difference $\Delta B_z$ for the coherent $ST_0$ operations, and the homogeneous magnetic field of 500 mT is applied along the z-axis. Radio frequency single-electron transistor (rf-set, red circle) discriminates the different charge states of the QD array. **b.** Stability diagrams of the 3 different double QDs. All $Q_1$, $Q_2$, and $Q_3$ are operated at the charge region (2,0) – (1,1) where n (m) denotes the electron number in the left (right) dot in the (n, m) notation. **c.** Correlated double sampling (CDS)

scheme for measuring the short-lived triplet signal. The gate signal is sampled for short period of time (~ 200 ns ) right after the manipulation pulse returns to the measurement point, and the baseline signal of the same width is subtracted from the gate signal to remove the background fluctuation, and averaged ~ 4000 times. The expected rf-signal when the operation pulse is applied is shown in green line in the upper panel. In the lower panel, the CDS stability diagram recorded concurrently with the $Q_2$ diagram in Fig. 1b is shown, where only the Pauli spin blockade region exhibits the low CDS voltage due to the superposed control pulse cycles. **d.** Relaxation time of the 3 $ST_0$ qubits. The time-averaged rf signal within the measurement windows are fitted to the exponential decay curve to yield the relaxation time $T_1$. The measured $T_1$ varies from $1\,\mu s$ to $11\,\mu s$ depending on the spatial $\Delta B_z$ distribution.

**Figure 2. Magnetic field simulation and coherent Larmor oscillations. a.** Coherent $\Delta B_z$ oscillation of the three $ST_0$ qubits. $\Delta B_z$ oscillation frequency 76 MHz ($Q_1$), 311 MHz ($Q_2$), and 192 MHz ($Q_3$) extracted from the fitting slightly differ from the numerical simulation which is expected to be due to micro-magnet misalignment. The oscillation curves are offset for clarity. **b.** Numerical simulation of the micro-magnet field calculated by boundary integral method using RADIA[46,47]. The dots denote the expected position of the QDs. In the lower panel, line-cut of the simulation at z = 0 is shown, and the expected (measured) $\Delta B_z$ at the qubit sites are shown in green void (solid) dots.

**Figure 3. Two axis control of the $ST_0$ qubits and noise analysis. a.** Free induction decay (FID) of $Q_1$, $Q_2$, and $Q_3$ measured by the $(\pi/2)_x - J(\varepsilon) - (\pi/2)_x$ pulse sequence (upper panel of the inset). FID curves at different $A_{ex}$ are shown in the waterfall plot ($Q_1$ – orange, $Q_2$ –

green, Q$_3$ – blue), where increasing the $|A_{ex}|$ decreases the J($\varepsilon$) resulting in smaller oscillation frequency as evident from the energy diagram in the lower panel of the inset. **b.** Dependence of the oscillation frequency on $A_{ex}$. Empirically assuming the exponential dependence of the J($\varepsilon$) on the $\varepsilon$ [21,50], the oscillation frequency follows $f(\varepsilon) = \sqrt{\Delta B_z^2 + J(\varepsilon)^2}/h$. **c.** Noise analysis of the three qubits. The root-mean-squared (rms) noise amplitudes can be extracted from the fitting of the $T_2^*$ vs $E_{tot}$ plot, which yields rms charge noise of ($6.3 \pm 0.6\,\mu\text{eV}$, $5.3 \pm 0.4\,\mu\text{eV}$, $9.5 \pm 2.5\,\mu\text{eV}$), and magnetic field noise of ($52.0 \pm 3.0\,\text{neV}$, $39.2 \pm 2.2\,\text{neV}$, $50.1 \pm 5.79\,\text{neV}$), for (Q$_1$, Q$_2$, Q$_3$).

# References

[1] A.J. Sigillito, J.C. Loy, D.M. Zajac, M.J. Gullans, L.F. Edge, and J.R. Petta, Phys. Rev. Applied **11**, 061006 (2019).

[2] D.M. Zajac, T.M. Hazard, X. Mi, E. Nielsen, and J.R. Petta, Phys. Rev. Applied **6**, 054013 (2016).

[3] C. Volk, A.M.J. Zwerver, U. Mukhopadhyay, P.T. Eendebak, C.J. van Diepen, J.P. Dehollain, T. Hensgens, T. Fujita, C. Reichl, W. Wegscheider, and L.M.K. Vandersypen, Npj Quantum Inf. **5**, 29 (2019).

[4] T. Ito, T. Otsuka, T. Nakajima, M.R. Delbecq, S. Amaha, J. Yoneda, K. Takeda, A. Noiri, G. Allison, A. Ludwig, A.D. Wieck, and S. Tarucha, Appl. Phys. Lett. **113**, 093102 (2018).

[5] J.R. Petta, A.C. Johnson, J.M. Taylor, E.A. Laird, A. Yacoby, M.D. Lukin, C.M. Marcus, M.P. Hanson, and A.C. Gossard, Science **309**, 2180 (2005).

[6] M. Veldhorst, C.H. Yang, J.C.C. Hwang, W. Huang, J.P. Dehollain, J.T. Muhonen, S. Simmons, A. Laucht, F.E. Hudson, K.M. Itoh, A. Morello, and A.S. Dzurak, Nature **526**, 410 (2015).

[7] T.F. Watson, S.G.J. Philips, E. Kawakami, D.R. Ward, P. Scarlino, M. Veldhorst, D.E. Savage, M.G. Lagally, M. Friesen, S.N. Coppersmith, M.A. Eriksson, and L.M.K. Vandersypen, Nature **555**, 633 (2018).

[8] M.A. Fogarty, K.W. Chan, B. Hensen, W. Huang, T. Tanttu, C.H. Yang, A. Laucht, M. Veldhorst, F.E. Hudson, K.M. Itoh, D. Culcer, T.D. Ladd, A. Morello, and A.S. Dzurak, Nat. Commun. **9**, 4370 (2018).

[9] J. Yoneda, K. Takeda, T. Otsuka, T. Nakajima, M.R. Delbecq, G. Allison, T. Honda, T. Kodera, S. Oda, Y. Hoshi, N. Usami, K.M. Itoh, and S. Tarucha, Nat. Nanotechnol. **13**, 102 (2018).

[10] J.M. Nichol, L.A. Orona, S.P. Harvey, S. Fallahi, G.C. Gardner, M.J. Manfra, and A. Yacoby, Npj Quantum Inf. **3**, 3 (2017).

[11] M.D. Shulman, O.E. Dial, S.P. Harvey, H. Bluhm, V. Umansky, and A. Yacoby, Science **336**, 202 (2012).

[12] A. Morello, J.J. Pla, F.A. Zwanenburg, K.W. Chan, K.Y. Tan, H. Huebl, M. Möttönen, C.D. Nugroho, C. Yang, J.A. van Donkelaar, A.D.C. Alves, D.N. Jamieson, C.C. Escott, L.C.L. Hollenberg, R.G. Clark, and A.S. Dzurak, Nature **467**, 687 (2010).

[13] E. Kawakami, P. Scarlino, D.R. Ward, F.R. Braakman, D.E. Savage, M.G. Lagally, M. Friesen, S.N. Coppersmith, M.A. Eriksson, and L.M.K. Vandersypen, Nat. Nanotechnol. **9**, 666 (2014).

[14] F.H.L. Koppens, C. Buizert, K.J. Tielrooij, I.T. Vink, K.C. Nowack, T. Meunier, L.P. Kouwenhoven, and L.M.K. Vandersypen, Nature **442**, 766 (2006).

[15] J.P. Dodson, N. Holman, B. Thorgrimsson, S.F. Neyens, E.R. MacQuarrie, T. McJunkin, R.H. Foote, L.F. Edge, S.N. Coppersmith, and M.A. Eriksson, ArXiv:2004.05683 [Cond-Mat, Physics:Physics, Physics:Quant-Ph] (2020).

[16] L.P. Kouwenhoven, N.C. van der Vaart, A.T. Johnson, W. Kool, C.J.P.M. Harmans, J.G. Williamson, A.A.M. Staring, and C.T. Foxon, Z. Physik B - Condensed Matter **85**, 367 (1991).

[17] D. Loss and D.P. DiVincenzo, Phys. Rev. A **57**, 120 (1998).

[18] M. Veldhorst, J.C.C. Hwang, C.H. Yang, A.W. Leenstra, B. de Ronde, J.P. Dehollain, J.T. Muhonen, F.E. Hudson, K.M. Itoh, A. Morello, and A.S. Dzurak, Nat. Nanotechnol. **9**, 981 (2014).

[19] S. Foletti, H. Bluhm, D. Mahalu, V. Umansky, and A. Yacoby, Nat. Phys. **5**, 903 (2009).

[20] B.M. Maune, M.G. Borselli, B. Huang, T.D. Ladd, P.W. Deelman, K.S. Holabird, A.A. Kiselev, I. Alvarado-Rodriguez, R.S. Ross, A.E. Schmitz, M. Sokolich, C.A. Watson, M.F. Gyure, and A.T. Hunter, Nature **481**, 344 (2012).

[21] X. Wu, D.R. Ward, J.R. Prance, D. Kim, J.K. Gamble, R.T. Mohr, Z. Shi, D.E. Savage, M.G. Lagally, M. Friesen, S.N. Coppersmith, and M.A. Eriksson, Proceedings of the National Academy of Sciences **111**, 11938 (2014).

[22] E.A. Laird, J.M. Taylor, D.P. DiVincenzo, C.M. Marcus, M.P. Hanson, and A.C. Gossard, Phys. Rev. B **82**, 075403 (2010).

[23] J. Medford, J. Beil, J.M. Taylor, S.D. Bartlett, A.C. Doherty, E.I. Rashba, D.P. DiVincenzo, H. Lu, A.C. Gossard, and C.M. Marcus, Nature Nanotech **8**, 654 (2013).

[24] J. Medford, J. Beil, J.M. Taylor, E.I. Rashba, H. Lu, A.C. Gossard, and C.M. Marcus, Phys. Rev. Lett. **111**, 050501 (2013).

[25] A.R. Mills, D.M. Zajac, M.J. Gullans, F.J. Schupp, T.M. Hazard, and J.R. Petta, Nat. Commun. **10**, 1063 (2019).

[26] T. Hensgens, T. Fujita, L. Janssen, X. Li, C.J. Van Diepen, C. Reichl, W. Wegscheider, S. Das Sarma, and L.M.K. Vandersypen, Nature **548**, 70 (2017).

[27] T.-K. Hsiao, C.J. van Diepen, U. Mukhopadhyay, C. Reichl, W. Wegscheider, and L.M.K. Vandersypen, Phys. Rev. Applied **13**, 054018 (2020).

[28] H. Bluhm, S. Foletti, D. Mahalu, V. Umansky, and A. Yacoby, Phys. Rev. Lett. **105**, 216803 (2010).

[29] M. Seo, H.K. Choi, S.-Y. Lee, N. Kim, Y. Chung, H.-S. Sim, V. Umansky, and D. Mahalu, Phys. Rev. Lett. **110**, 046803 (2013).

[30] C. Hong, G. Yoo, J. Park, M.-K. Cho, Y. Chung, H.-S. Sim, D. Kim, H. Choi, V. Umansky, and D. Mahalu, Phys. Rev. B **97**, 241115 (2018).

[31] T. Ito, T. Otsuka, T. Nakajima, M.R. Delbecq, S. Amaha, J. Yoneda, K. Takeda, A. Noiri, G. Allison, A. Ludwig, A.D. Wieck, and S. Tarucha, Appl. Phys. Lett. **113**, 093102 (2018).

[32] J. Yoneda, T. Otsuka, T. Takakura, M. Pioro-Ladrière, R. Brunner, H. Lu, T. Nakajima, T. Obata, A. Noiri, C.J. Palmstrøm, A.C. Gossard, and S. Tarucha, Appl. Phys. Express **8**, 084401 (2015).

[33] D. Maradan, L. Casparis, T.-M. Liu, D.E.F. Biesinger, C.P. Scheller, D.M. Zumbühl, J.D. Zimmerman, and A.C. Gossard, J Low Temp Phys **175**, 784 (2014).

[34] C. Livermore, C.H. Crouch, R.M. Westervelt, K.L. Campman, and A.C. Gossard, Science **274**, 1332 (1996).

[35] W. Jang, J. Kim, M.-K. Cho, H. Chung, S. Park, J. Eom, V. Umansky, Y. Chung, and D. Kim, Npj Quantum Inf **6**, 64 (2020).

[36] W. Jang, M.-K. Cho, M. Lee, C. Hong, J. Kim, H. Jung, Y. Chung, V. Umansky, and D. Kim, Appl. Phys. Lett. **114**, 242102 (2019).

[37] J. Stehlik, Y.-Y. Liu, C.M. Quintana, C. Eichler, T.R. Hartke, and J.R. Petta, Phys. Rev. Applied **4**, 014018 (2015).

[38] T. Nakajima, M.R. Delbecq, T. Otsuka, P. Stano, S. Amaha, J. Yoneda, A. Noiri, K. Kawasaki, K. Takeda, G. Allison, A. Ludwig, A.D. Wieck, D. Loss, and S. Tarucha, Phys. Rev. Lett. **119**, 017701 (2017).

[39] C. Barthel, J. Medford, H. Bluhm, A. Yacoby, C.M. Marcus, M.P. Hanson, and A.C. Gossard, Phys. Rev. B **85**, 035306 (2012).

[40] L.A. Orona, J.M. Nichol, S.P. Harvey, C.G.L. Bøttcher, S. Fallahi, G.C. Gardner, M.J. Manfra, and A. Yacoby, Phys. Rev. B **98**, 125404 (2018).

[41] K.D. Petersson, J.R. Petta, H. Lu, and A.C. Gossard, Phys. Rev. Lett. **105**, 246804 (2010).

[42] D. Kim, D.R. Ward, C.B. Simmons, J.K. Gamble, R. Blume-Kohout, E. Nielsen, D.E. Savage, M.G. Lagally, M. Friesen, S.N. Coppersmith, and M.A. Eriksson, Nature Nanotechnology **10**, 243 (2015).

[43] C. Barthel, D.J. Reilly, C.M. Marcus, M.P. Hanson, and A.C. Gossard, Phys. Rev. Lett. **103**, 160503 (2009).

[44] P. Harvey-Collard, B. D'Anjou, M. Rudolph, N.T. Jacobson, J. Dominguez, G.A. Ten Eyck, J.R. Wendt, T. Pluym, M.P. Lilly, W.A. Coish, M. Pioro-Ladrière, and M.S. Carroll, Phys. Rev. X **8**, 021046 (2018).

[45] D.J. Reilly, J.M. Taylor, J.R. Petta, C.M. Marcus, M.P. Hanson, and A.C. Gossard, Science **321**, 817 (2008).

[46] P. Elleaume, O. Chubar, and J. Chavanne, in *Proceedings of the 1997 Particle Accelerator Conference (Cat. No.97CH36167)* (1997), pp. 3509–3511 vol.3.

[47] O. Chubar, P. Elleaume, and J. Chavanne, J. Synchrotron Rad. **5**, 481 (1998).

[48] M.D. Shulman, S.P. Harvey, J.M. Nichol, S.D. Bartlett, A.C. Doherty, V. Umansky, and A. Yacoby, Nat. Commun. **5**, 5156 (2014).

[49] T. Nakajima, A. Noiri, K. Kawasaki, J. Yoneda, P. Stano, S. Amaha, T. Otsuka, K. Takeda, M.R. Delbecq, G. Allison, A. Ludwig, A.D. Wieck, D. Loss, and S. Tarucha, Phys. Rev. X **10**, 011060 (2020).

[50] O.E. Dial, M.D. Shulman, S.P. Harvey, H. Bluhm, V. Umansky, and A. Yacoby, Phys. Rev. Lett. **110**, 146804 (2013).

[51] A. Pateras, J. Park, Y. Ahn, J.A. Tilka, M.V. Holt, C. Reichl, W. Wegscheider, T.A. Baart, J.P. Dehollain,

U. Mukhopadhyay, L.M.K. Vandersypen, and P.G. Evans, Nano Lett. **18**, 2780 (2018).

[52] Yong Cai, Yugang Zhou, K.J. Chen, and K.M. Lau, IEEE Electron Device Letters **26**, 435 (2005).

[53] D.J. Reilly, C.M. Marcus, M.P. Hanson, and A.C. Gossard, Appl. Phys. Lett. **91**, 162101 (2007).

**Materials & Correspondence**

Correspondence and requests for materials should be addressed to D.K. (dohunkim@snu.ac.kr ).

**Supplementary Information**

**S1. Carrier depletion under micro-magnet**

By performing transport measurements through the ohmic contacts around the single electron transistor sensors (rf-set, Supplementary Fig. S1a), we find the evidence for the two-dimensional electron gas (2DEG) depletion under the micro-magnet region. For example, we examined the effect of the right rf-set sensor gates ($B1_{RS}$, $P_{RS}$, and $B2_{RS}$) to the left sensor current $I_{Oc-OL}$ while left sensor quantum dot is formed. As shown in Fig. 1b, the current which is supposed to flow from the center to left Ohmic ($O_c$-$O_L$) can be completely pinched off by right sensor gates reflecting that the actual current path necessarily includes right sensor region (see inset of Supplementary Fig. S1b for reasonable guess for the current path). After examining more than 10 identically fabricated devices, we find that the effect reproducibly occurs, and we conclude that the area directly under surface-deposited micro-magnet is carrier-free likely due to interface strain[1] or the plasma treatment[2] during the fabrication process.

While this adverse effect prevents simultaneous forming of two rf-set sensors to the maximally sensitive position, independent rf-sensor can be tuned and good rf-matching is possible by grounding the other sensor gates thereby securing proper rf current path as shown in Supplementary Fig. 1c and 1d. Thus, we operate only one rf-set sensor in the current work and plan to add gating capability to the micro-magnet by connecting it to the additional gate electrode in the future.

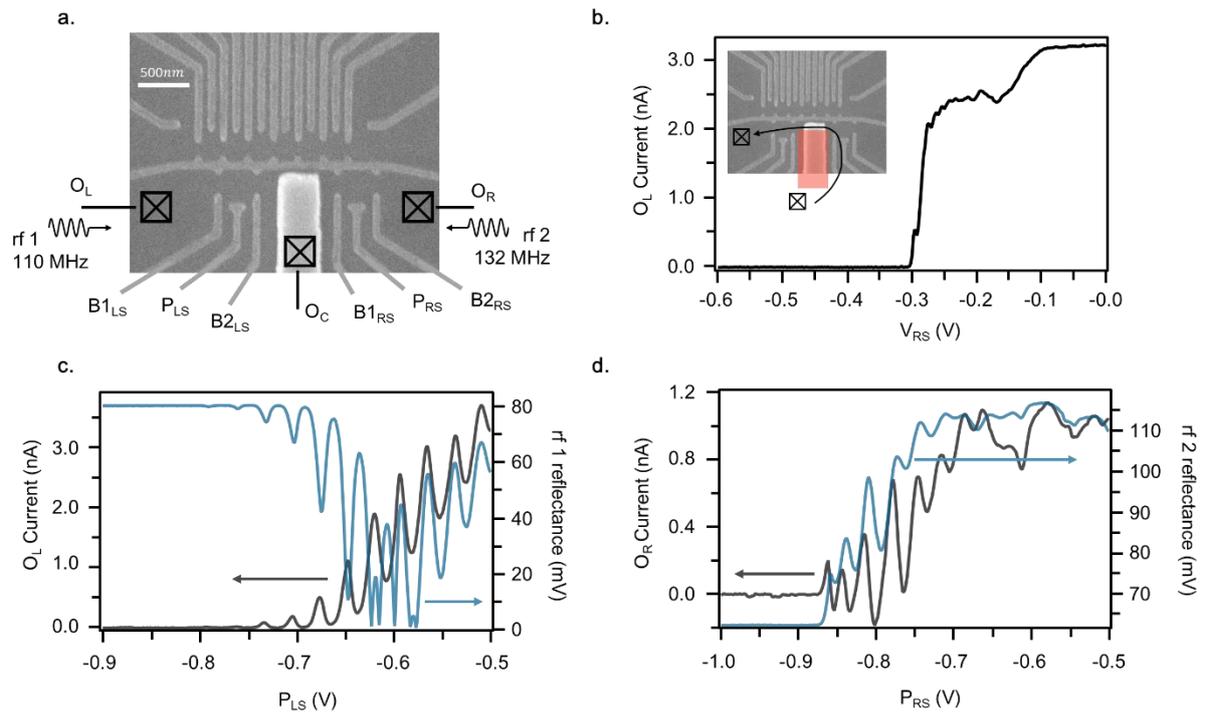

**Supplementary Figure S4**. **Calibration of rf single electron transistor sensors (rf-set), and evidence for two-dimensional electron gas (2DEG) depletion under micro-magnet. a.** Scanning electron microscope (SEM) image of the device similar to the device used in the experiment. Three ohmic contacts ($O_L$, $O_C$ and $O_R$) can be utilized to directly observe the conductance change through the rf-set channels. Two different high-frequency lines are installed to $O_L$ and $O_R$ respectively for rf-reflectometry measurements. **b.** Current from center Ohmic $O_C$ to left Ohmic $O_L$ as a function of gate voltage simultaneously applied to right sensor gates ($B1_{RS}$, $P_{RS}$, and $B2_{RS}$) while left sensor quantum dot is formed. **c.** (**d.**) Coulomb oscillation of the rf-set1 (2). The direct current through $O_L$($O_R$), and the rf 1(2) reflectance were concurrently recorded as a function of plunger gate voltage $P_{LS}$ ($P_{RS}$) while the three rf-set2 (1) gates were grounded, showing capability of good independent sensor matching.

## Supplementary References

[1] A. Pateras, J. Park, Y. Ahn, J.A. Tilka, M.V. Holt, C. Reichl, W. Wegscheider, T.A. Baart, J.P. Dehollain, U. Mukhopadhyay, L.M.K. Vandersypen, and P.G. Evans, Nano Lett. **18**, 2780 (2018).
[2] Yong Cai, Yugang Zhou, K.J. Chen, and K.M. Lau, IEEE Electron Device Letters **26**, 435 (2005).